%% file: arrival_time_UC_POVM.tex
\documentclass[11pt]{article}

\usepackage{amsmath,amssymb,amsthm}
\usepackage{bm}
\usepackage{geometry}
\usepackage{graphicx}
\usepackage[most]{tcolorbox}
\usepackage{hyperref}

\title{The Unidirectional Current as First Arrival-Time POVM: An MS--Kijowski Identity, Physical Interpretation, and Mathematical Applications}
\author{
Avi Marchewka\\ 
\small 8 Galei Tchelet Street, Herzliya, Israel\\
\small\href{mailto:Avi.marchewka@gmail.com}{Avi.marchewka@gmail.com}
}
\date{}

\usepackage{cite}

\begin{document}

\maketitle

%%%%%%%%%%%%%%%%%%%%%%%%%%%%%%%%%%%%%%%%%%%%%%%%%%%%%%

\begin{abstract}
\input{sections/abstract}

\end{abstract}
%%%%%%%%%%%%%%%%%%%%%%%%%%%%%%%%%%%%%%%%%%%%%%%%%%%%%%

%%%%%%%%%%%%%%%%%%%%%%%%%%%%%%%%%%%%%%%%%%%%%%%%%%%%%%
\input{sections/introduction}
 \section{Original Representations of the MS Unidirectional First-Arrival Current}
\label{sec:two_MS_representations}

The MS construction is formulated within the Feynman path-integral description of quantum propagation. In this framework, the unidirectional first-arrival current is constructed by classifying the Feynman paths contributing to the propagation amplitude according to their relation to the detector boundary. Accordingly, throughout this work, whenever we refer to a path or trajectory in the MS construction, we mean a Feynman path contributing to the quantum propagation amplitude.

The purpose of this section is to provide a brief and structured presentation of the elements and concepts underlying the MS unidirectional-current model \cite{MarchewkaSchuss1998,MarchewkaSchuss2000,MarchewkaSchuss2002} that will be used in the present work. We proceed in the following order. First, we construct the unidirectional current from the Dirichlet boundary, the first-arrival condition, and the length parameter \(\lambda\). We then introduce the statistical interpretation of this current as a hazard rate.

\label{subsec:original_MS_representation}

\paragraph{Dirichlet Boundary, First Arrival, and the Unidirectional Current.}

\begin{figure*}[t]
\centering
\includegraphics[width=\textwidth]{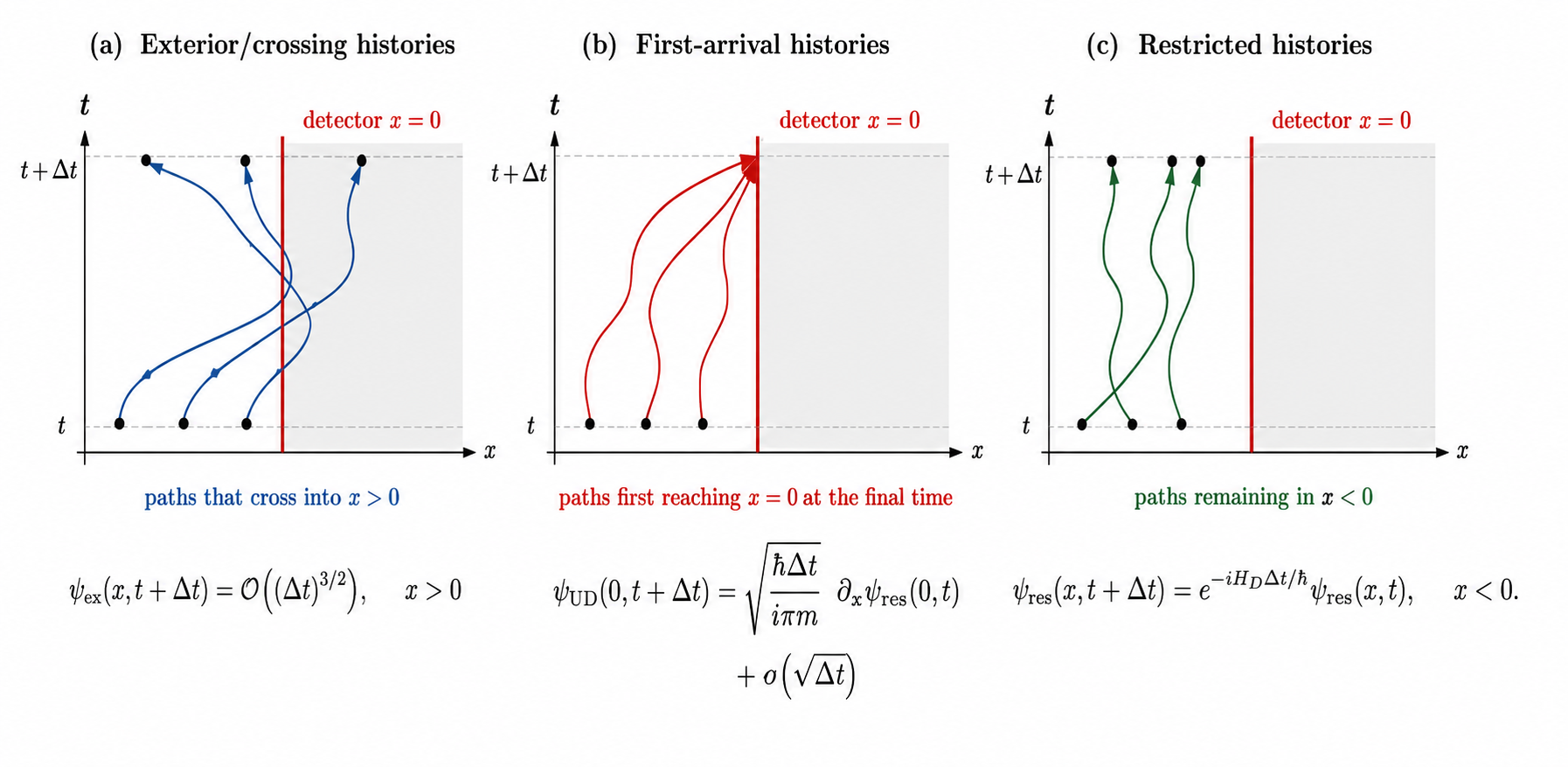}
\caption{Path decomposition relative to the detector boundary at $x=0$.
(a) Histories that reach the boundary and propagate into the exterior region.
(b) Histories whose first arrival at the boundary occurs at $t+\Delta t$.
(c) Restricted histories that remain in $x<0$ throughout the interval
$[t,t+\Delta t]$.}
\label{fig:path_decomposition}
\end{figure*}

We place the detector at the origin, \(x_D=0\), and assume that the
particle is initially confined to the negative half-line. Relative to the detector boundary, the Feynman paths contributing to the short-time propagation amplitude over an interval \(\Delta t\) can be divided into three sectors; see Fig.~\ref{fig:path_decomposition}. The first consists of restricted trajectories
that remain in the region \(x<0\) and do not reach the detector. The
second consists of trajectories that reach the boundary \(x=0\) for
the first time during the interval. The third consists of trajectories
that cross into the exterior region \(x>0\) at least once.

In the short-time limit, the corresponding propagation can be written
as
\begin{equation}
    \label{eq:prop}
\psi(x,t)
\;\xrightarrow{\;\Delta t\;}\;
\begin{cases}
\displaystyle
\psi_{\mathrm{ex}}(x,t+\Delta t)
    =O\!\left((\Delta t)^{3/2}\right),
& x>0,
\\begin{equation}8pt]
\displaystyle
\psi_{\mathrm{UD}}(0,t+\Delta t)
    =
    \sqrt{\frac{\hbar\Delta t}{i\pi m}}\,
    \partial_x\psi_{\mathrm{res}}(0,t)
    +o\!\left(\sqrt{\Delta t}\right),
& x=0,
\\begin{equation}10pt]
\displaystyle
\psi_{\mathrm{res}}(x,t+\Delta t)
    =
    e^{-iH_D\Delta t/\hbar}
    \psi_{\mathrm{res}}(x,t),
& x<0.
\end{cases}
\end{equation}
Here, \(H_D\) is the Hamiltonian generating the free propagation on
the negative half-line, subject to the Dirichlet boundary condition
\begin{equation*}
\psi_{\mathrm{res}}(0,t)=0.
\end{equation*}
The restricted wave function \(\psi_{\mathrm{res}}(x,t)\) describes
the particle conditioned on not having reached the detector before
time \(t\). In the notation used here, it is normalized at each time,
\begin{equation*}
\left\lVert\psi_{\mathrm{res}}(t)\right\rVert^2=1.
\end{equation*}

To describe the probability transferred into each sector during the
short interval, we define the local probability-density rate by
\begin{equation}
\mathcal{J}(x,t^+)
=
\lim_{\Delta t\to0}
\frac{
\rho(x,t+\Delta t)-\rho(x,t)
}{
\Delta t
},
\qquad
\rho(x,t)=|\psi(x,t)|^2.
\end{equation}
For the boundary and exterior sectors, the initial density in this
expression vanishes. The three sectors therefore give
 \begin{equation}
\mathcal{J}(x,t^+)=
\begin{cases}
\displaystyle
\mathcal{J}_{\rm ex}(x,t)=0,
& x>0,\\[8pt]

\displaystyle
\mathcal{J}_{\rm UD}(0,t)
=
\frac{\hbar}{\pi m}
\left|\partial_x\psi_{\rm res}(0,t)\right|^2,
& x=0,\\[10pt]

\displaystyle
-\partial_x j_{\rm Sch}(x,t),
\qquad
j_{\rm Sch}(x,t)
=
\frac{\hbar}{m}\operatorname{Im}
[\psi_{\rm res}^*(x,t)\partial_x\psi_{\rm res}(x,t)],
& x<0 .
\end{cases}
\label{eq:boundary_current}
\end{equation}
Thus, within the restricted half-line, the probability-density rate
is governed by the ordinary Schrödinger continuity equation. At the
boundary, the short-time propagator produces the positive
unidirectional first-arrival density rate \(j_{\mathrm{UD}}(0,t)\),
whereas the corresponding rate in the exterior region vanishes in
the limit \(\Delta t\to0\).

The next step in the MS construction is to model detection as the
absorption of the first-arrival contribution when it reaches the
boundary. If no detection event occurs, the particle has not reached the detector, and its quantum evolution is described by the restricted Feynman paths that do not cross the detector boundary. The same construction is then repeated during the next time interval.

The boundary quantity \(j_{\mathrm{UD}}(0,t)\) has the dimensions of
probability per unit length per unit time. A phenomenological
detection parameter \(\lambda\), with dimensions of length, is
therefore introduced to convert it into the unidirectional detection
current
\begin{equation}
J_{\mathrm{UD}}(t)
=
\lambda\,j_{\mathrm{UD}}(0,t)
=
\lambda\frac{\hbar}{\pi m}
\left|
\partial_x\psi_{\mathrm{res}}(0,t)
\right|^2.
\label{eq:MS_unidirectional_current}
\end{equation}
The resulting quantity \(J_{\mathrm{UD}}(t)\) has dimensions of an
inverse time and represents a probability rate. Accordingly,
\(J_{\mathrm{UD}}(t)\,dt\) gives the conditional probability that the
particle first reaches the boundary during the interval
\([t,t+dt]\), given that it has not reached it earlier. In the MS
picture, first arrival at the boundary is identified with detection.
The unidirectional current is therefore the central quantity connecting
the restricted quantum dynamics with the first- and last-arrival-time
statistics. 

At this stage, \(\lambda\) is introduced as an ad hoc phenomenological
length parameter. Its role is to relate the first-arrival density
rate at the boundary to the current representing the detection
process. In Sec.~\ref{sec:lamda structure}, we return to this
parameter and examine its manifestation at the wave-function level.

 \paragraph{Conditional-Hazard Interpretation.}

A common interpretation of the MS construction is to regard the
unidirectional current as a conditional first-arrival hazard rate,
\begin{equation}
h_{\mathrm{MS}}(t)
=
J_{\mathrm{UD}}[\psi_{\mathrm{res}}(t)].
\label{eq:MS_hazard}
\end{equation}

Under this interpretation,
\begin{equation*}
\Pr\!\left(
t<T_{\mathrm{arr}}\leq t+dt
\,\middle|\,
T_{\mathrm{arr}}>t
\right)
=
h_{\mathrm{MS}}(t)\,dt+o(dt),
\label{eq:MS_conditional_arrival}
\end{equation*}
where \(\psi_{\mathrm{res}}(t)\) is understood as the normalized state conditioned on the absence of an earlier arrival.

The corresponding survival probability is then constructed as
\begin{equation*}
S_{\mathrm{MS}}(t)
=
\exp\!\left[
-\int_0^t h_{\mathrm{MS}}(\tau)\,d\tau
\right],
\label{eq:MS_survival_probability}
\end{equation*}
and the associated unconditional first-arrival density is
\begin{equation}
p_{\mathrm{MS}}(t)
=
-S_{\mathrm{MS}}'(t)
=
S_{\mathrm{MS}}(t)\,h_{\mathrm{MS}}(t).
\label{eq:MS_unconditional_density}
\end{equation}

Equivalently, one may introduce the unnormalized surviving state
\begin{align*}
\Psi_{\mathrm{surv}}(x,t)
&=
\sqrt{S_{\mathrm{MS}}(t)}\,\psi_{\mathrm{res}}(x,t),
\nonumber\\
\left\lVert\Psi_{\mathrm{surv}}(t)\right\rVert^2
&=
S_{\mathrm{MS}}(t).
\label{eq:MS_surviving_state}
\end{align*}
This conditional-hazard formulation provides a useful way of relating the instantaneous MS cur-
rent to the accumulated first-arrival statistics. We present this interpretation because it has been
adopted in parts of the subsequent literature and provides one possible statistical interpretation of
the MS current. In the present work, we consider a different possibility: interpreting the unidirec-
tional current through a positive operator and the generalized Born rule. .

\input{sections/positive_operator}

\input{sections/free_normalization}

\input{sections/statistical_interpretations}

\section{An MS--Kijowski Identity: Physical Interpretation and Mathematical Applications}
\label{sec:kijowski_ms}

An interesting structural observation follows from the uniquely normalized
free-particle response. The spectral form of the one-sided
unidirectional-current amplitude coincides with that of one directional
branch of Kijowski's arrival-time distribution
\cite{Kijowski1974,MugaSalaPalao1998,DasNoth2021,Naidon2024}. We establish
this relation below and then clarify its scope.

 For a particle approaching the detector at \(x=0\) from the left, the MS
state belongs to \(L^2(\mathbb R_-)\), satisfies the Dirichlet condition at
the origin, and has the spectral expansion
\begin{equation*}
 \psi_L(x)
 =
 \int_0^\infty \widetilde\psi_L(k)\,
 \phi_k^L(x)\,dk ,
 \qquad
 \phi_k^L(x)=\sqrt{\frac{2}{\pi}}\sin(kx).
\end{equation*}

Here \(k\geq0\) labels the continuous Dirichlet spectrum, with
\(k=0\) understood as the spectral threshold. Accordingly, the spectral amplitude
\(\widetilde\psi_L(k)\) is defined only for \(k\geq0\). This should not be
interpreted as a restriction of the physical state to positive plane-wave
momentum. Indeed, each sine mode contains both wave-number signs,
\begin{equation*}
 \phi_k^L(x)
 =
 \frac{1}{i\sqrt{2\pi}}
 \left(e^{ikx}-e^{-ikx}\right).
 \label{eq:left_dirichlet_modes}
\end{equation*}

POVM normalization of the unidirectional current uniquely fixes the
free-particle response \(\lambda(k)\) of the boundary amplitude,
Eq.~(\ref{eq:Lamda normalized}), giving
\begin{equation*}
 B_{L,t}^{(\lambda)}\psi_L
 =
 i\sqrt{\frac{\hbar}{2\pi m}}
 \int_0^\infty
 \sqrt{k}\,\widetilde\psi_L(k)
 e^{-i\hbar k^2t/(2m)}\,dk .
\end{equation*}

The spectral structure of this expression coincides with that of
Kijowski's positive-momentum amplitude,
\begin{equation*}
 A_+(t)
 =
 \sqrt{\frac{\hbar}{2\pi m}}
 \int_0^\infty
 \sqrt{k}\,\widetilde\psi_+(k)
 e^{-i\hbar k^2t/(2m)}\,dk .
\end{equation*}

To relate the two representations, we use the normalized full-line
extension of the half-line state. Up to an irrelevant phase convention,
the positive-momentum component of the extended state satisfies
\begin{equation*}
 \widetilde\psi_+(k)
 =
 \frac{1}{\sqrt{2}}\widetilde\psi_L(k),
 \qquad k>0 .
\end{equation*}

Consequently,

\begin{equation}
 B_{L,t}^{(\lambda)}\psi_L
 =
 i\sqrt{2}\,A_+(t),
 \qquad
 J_{\mathrm{UD},L}(t)
 =
 2|A_+(t)|^2 .
 \label{eq:left_equals_kijowski_positive}
\end{equation}

The same reasoning applies to a particle approaching the origin from the
right. One obtains

\begin{equation}
 B_{R,t}^{(\lambda)}\psi_R
 =
 i\sqrt{2}\,A_-(t),
 \qquad
 J_{\mathrm{UD},R}(t)
 =
 2|A_-(t)|^2 .
 \label{eq:left_equals_kijowski_Negativ}
\end{equation}

Thus each one-sided MS current reproduces the corresponding directional
branch of Kijowski's distribution, up to the normalization factor arising
from the comparison of the two representations.

Combining the two directional contributions gives

\begin{equation}
\Pi_K(t)
=
|A_+(t)|^2+|A_-(t)|^2
=
\frac{1}{2}
\left[
J_{\mathrm{UD},L}(t)+J_{\mathrm{UD},R}(t)
\right].
\label{eq:MS-Kijo_identetey}
\end{equation}

The factor \(1/2\) reflects the different normalization of the two
descriptions. Each one-sided MS current corresponds to a separately
normalized first-arrival process. Their direct sum therefore represents two
normalized directional processes. Kijowski's distribution, in contrast,
describes a single normalized arrival-time distribution. The relation above
corresponds to the symmetric normalization in which the two directional
processes are assigned equal weight.

\paragraph{Physical consequences of the MS--Kijowski identity.}

The equality in Eq.~\eqref{eq:left_equals_kijowski_positive} is an equality
of arrival-time statistics, not of coordinate-space wave functions or of
the underlying physical processes. In Kijowski's construction, \(A_+(t)\)
is evaluated for a freely propagating full-line state supported in the
positive-momentum sector. The associated POVM specifies an arrival-time
distribution, but by itself does not determine whether the registered event
represents a first crossing, a unique crossing, or a terminal detection.

The MS construction provides a different spatial realization of the same
temporal law. The state is confined to one side of the detector, the
Dirichlet boundary at \(x=0\) specifies the detector location, and the
unidirectional boundary current follows from the first-arrival construction.
Thus the particle approaches from a prescribed side, the registered event
is its first arrival at the boundary, and detection terminates the one-sided
process. The momentum content of the two descriptions is also different.
The positive- and negative-wave-number components appearing in the
plane-wave representation of the MS state are coherently related by the
Dirichlet condition and form a single spatial state. They therefore cannot
be interpreted as independent arrival alternatives. Thus, for a single
arrival direction, the MS construction and the corresponding Kijowski
branch give the same temporal statistics while representing different
spatial processes.

 The two directional Kijowski branches thus correspond to the two oppositely directed MS first-arrival processes.  
 To examine the physical meaning of this relation for a single particle, consider a normalized MS state distributed between the two one-sided arrival channels,

\begin{equation}
\mathcal H_{\mathrm{MS}}
=
\mathcal H_L\oplus\mathcal H_R,
\end{equation}
with
\begin{equation}
\Psi
=
\sqrt{w_L}\,\psi_L
\oplus
e^{i\phi}\sqrt{w_R}\,\psi_R,
\qquad
w_L+w_R=1.
\end{equation}
The corresponding unidirectional current is
\begin{equation}
J_{\mathrm{UD}}(t)
=
w_L J_{\mathrm{UD},L}(t)
+
w_R J_{\mathrm{UD},R}(t).
\label{eq:two_channel_MS}
\end{equation}
There is no interference term between the left- and right-arrival channels:
the two components represent spatially separated first-arrival processes on
opposite half-lines.

For the symmetric choice
\begin{equation}
w_L=w_R=\frac{1}{2},
\end{equation}
Eq.~\eqref{eq:two_channel_MS} becomes
\begin{equation}
J_{\mathrm{UD}}(t)
=
\frac{1}{2}
\left[
J_{\mathrm{UD},L}(t)+J_{\mathrm{UD},R}(t)
\right]
=
\Pi_K(t).
\end{equation}
Thus the factor \(1/2\) in the MS--Kijowski identity has a direct physical
interpretation in the MS realization: the normalized state is distributed
between two spatially separated first-arrival channels. Since these channels
form a direct sum, their contributions add without an interference term.
The identity is therefore exact at the level of arrival-time statistics,
while the underlying spatial realizations remain physically distinct.

 \paragraph{Mathematical consequences of the MS--Kijowski correspondence.}
The MS--Kijowski identity provides more than an equality between two
expressions for the arrival-time distribution. It allows properties of the
Kijowski distribution to be understood through the simpler spatial structure
of the MS representation. In the Kijowski formulation, both the long-time behavior of the
arrival-time distribution and the existence of its time moments are
determined by the low-momentum behavior of the spectral amplitude near
\(k=0\). The MS representation provides a direct way to analyze this regime
through the half-line Dirichlet structure. Independently of the correspondence, the unidirectional current itself
provides a framework for extracting the long-time asymptotic behavior
directly from the restricted dynamics.

For a particle arriving from the left, the MS spectral amplitude is given by
the sine-transform representation
\begin{equation*}
a(k)
=
\sqrt{\frac{2}{\pi}}
\int_{-\infty}^{0}\psi_L(x)\sin(kx)\,dx .
\end{equation*}
Because the MS state satisfies the Dirichlet boundary condition at the
detector,
\begin{equation*}
\psi_L(0)=0,
\end{equation*}
the low-\(k\) behavior follows directly from the expansion
\begin{equation*}
\sin(kx)\simeq kx ,
\qquad k\rightarrow0 .
\end{equation*}
For sufficiently regular states, this gives the generic behavior
\begin{equation*}
a(k)\sim k .
\end{equation*}
Thus the MS representation naturally explains the suppression of the
low-momentum contribution that determines the long-time tail.

More generally, if the MS spectral amplitude behaves as
\begin{equation*}
a(k)\sim k^\nu ,
\qquad k\rightarrow0 ,
\end{equation*}
then the MS--Kijowski identity implies that the corresponding Kijowski
arrival-time density has the asymptotic behavior
\begin{equation}
\Pi_K(t)\sim t^{-(\nu+3/2)} .
\end{equation}
In particular, the generic Dirichlet behavior
\begin{equation*}
a(k)\sim k
\end{equation*}
leads to
\begin{equation}
\Pi_K(t)\sim t^{-5/2}.
\end{equation}

This asymptotic behavior determines the existence of the arrival-time
moments,
\begin{equation}
\left\langle T^n\right\rangle
=
\int_0^\infty t^n\Pi_K(t)\,dt .
\end{equation}
The \(n\)-th moment is finite whenever
\begin{equation}
\nu>n-\frac12 .
\end{equation}
Therefore, if \(a(k)\) approaches a nonzero constant at \(k=0\), the
corresponding \(t^{-3/2}\) tail leads to a divergent mean arrival time.
For \(a(k)\sim k\), the tail becomes \(t^{-5/2}\), so that
\(\langle T\rangle\) is finite while
\(\langle T^2\rangle\) diverges. For \(a(k)\sim k^2\), the tail is
\(t^{-7/2}\), and both the first and second moments are finite. It is important to emphasize that this behavior has a different
interpretation in the two formulations. Although the Dirichlet condition
constitutes a mathematical restriction, it has a direct physical
interpretation as a detector boundary condition. Consequently, the
low-momentum behavior responsible for the long-time decay emerges naturally
from the restricted dynamics in the MS representation. In contrast, in the
Kijowski formulation the corresponding low-momentum behavior is a property
of the assumed initial momentum distribution.

The long-time tail therefore provides a possible experimental signature of
different arrival-time models. It has been proposed that different arrival-time models may
lead to distinct long-time decay exponents, making the tail behavior an
experimentally testable signature for distinguishing between competing
descriptions of quantum detection \cite{RafsanjaniCavendish2026}.

This viewpoint is complementary to the analysis of Das and Nöth
\cite{DasNoth2021}, who showed that the low-momentum behavior of the
Kijowski--Aharonov--Bohm state determines the existence of arrival-time
moments. The MS representation provides a spatial interpretation of this
low-momentum structure: the behavior near \(k=0\) follows from the
half-line Dirichlet realization and is then transferred to the Kijowski
distribution through the MS--Kijowski correspondence.

\input{sections/discussion}

 \appendix
\setcounter{section}{0}
\input{sections/appendix_kijowski}

\bibliographystyle{unsrt}
\bibliography{references_revised}
\end{document}

%% file: sections/abstract.tex
Detector-based first-arrival models and operator-based arrival-time observables provide two conceptually distinct approaches to the quantum time-of-arrival problem. Here we connect these approaches by using the positive unidirectional first-arrival current of Marchewka and Schuss (MS) as the basis for constructing a family of positive arrival-time operators and, upon normalization, an arrival-time POVM. The first-arrival character is inherited from the dynamics with a Dirichlet boundary, while the non-negative MS detection parameter, $\lambda$, is implemented at the amplitude level through a spectral response $\lambda(k)$.

For a one-dimensional free particle, resolution of the identity uniquely selects the spectral response
\[
\lambda(k)=\frac{\pi}{4k},
\]
thereby yielding an arrival-time POVM. The resulting normalized unidirectional current can therefore be used directly as a generalized Born- rule operator density rather than as a hazard rate, as in the original MS formulation. Here, by contrast, the normalized current itself defines the arrival-time probability density through a POVM.

As a secondary result, the normalized one-sided MS amplitude coincides, up to an irrelevant phase, with the corresponding momentum-sector amplitude in Kijowski's arrival-time distribution. Consequently, the two contributions to Kijowski's distribution can be reproduced by two first-arrival problems defined on opposite sides of the boundary. This establishes a mathematical equivalence between the corresponding temporal densities, although the underlying physical pictures remain fundamentally different: MS describes a first-arrival process and allows coherent interference between momentum components, whereas Kijowski's arrival-time POVM treats the two momentum sectors as separate directional contributions and contains no interference term between them.

Thus, under the generalized Born- rule interpretation introduced here, the normalized unidirectional current defines a POVM describing the particle's first and only arrival at a point.

%% file: sections/introduction.tex
\section{Introduction}

The quantum time-of-arrival (TOA) problem has remained a longstanding
conceptual challenge since the pioneering work of Aharonov and Bohm
\cite{AharonovBohm1961} and the systematic analysis of Allcock
\cite{Allcock1969}. Unlike position or momentum,
arrival time is not represented by a universally accepted quantum observable.
In particular, Pauli's objection excludes a self-adjoint time operator
canonically conjugate to a semibounded Hamiltonian \cite{Pauli1980}. This has
motivated several alternative formulations
\cite{Kijowski1974,Werner1987,MugaLeavens2000,Echanobe2008ZenoArrival}, while
recent theoretical and experimental proposals continue to address both the
foundational and operational aspects of the problem
\cite{Tumulka2022AnnPhys,GoldsteinTumulkaZanghi2024ArrivalDetection,
NaidonHappBoiron2026Proposal,RafsanjaniCavendish2026}.

Existing approaches may be grouped broadly into three classes. The first
constructs arrival-time observables, most notably positive operator-valued
measures (POVMs), of which Kijowski's distribution is the canonical
free-particle example \cite{Kijowski1974,Werner1987}. The second models the
measurement process through absorbing boundaries, absorbing potentials, or
related detector dynamics
\cite{MarchewkaSchuss1998,MarchewkaSchuss2000,MarchewkaSchuss2002,
Halliwell2008,Tumulka2023ABC}. A third class employs semiclassical or
probability-current constructions under appropriate physical conditions
\cite{Leavens1993,MugaLeavens2000}. These approaches differ not only in
mathematical form but also in what they identify as the physical arrival
event.

In the present work, we propose an alternative interpretation of the original MS unidirectional current, in which the current is used to define a POVM and the corresponding arrival-time probability through a generalized Born rule. This proposal is distinct from the original MS interpretation of the same current as a conditional first-arrival hazard. We do not regard the latter interpretation as incorrect; rather, the two constitute different probabilistic descriptions built upon the same underlying unidirectional current, whose physical applicability is ultimately an empirical question.

The remainder of the paper is organized as follows. Section (\ref{sec:two_MS_representations}) reviews the central elements of the MS proposal \cite{MarchewkaSchuss1998,MarchewkaSchuss2000, MarchewkaSchuss2002}, its statistical interpretation, and the quantities needed in the present work. Section(\ref{sec:lamda structure}) analyzes the manifestation of the free parameter $\lambda$ appearing in the MS model at the wave-function level, leading to its momentum-dependent amplitude representation $\lambda(k)$. Section~(\ref{sec:free_normalization}) constructs the corresponding family of positive operators, introduces the generalized Born-rule interpretation of the arrival-time law, and shows how resolution of the identity uniquely determines the normalized spectral response.
Section~(\ref{sec:kijowski_ms}) establishes, for a free particle, the correspondence between the normalized one-sided construction and Kijowski’s directional arrival-time amplitudes. Section~(\ref{sec:conclusion}) summarizes the main results and discusses their physical interpretation and implications.

%% file: sections/positive_operator.tex
\section{The Structure of $\lambda(k)$ at the Wave-Function Level}
\label{sec:lamda structure}
So far, the parameter \(\lambda\) has been introduced phenomenologically,
on dimensional grounds, as a multiplicative factor at the level of the
current. We now consider a more general formulation. First, we introduce
the response at the level of the wave function, through a factor
\(\sqrt{\lambda}\), so that the corresponding current remains proportional
to \(\lambda\). This formulation also provides a natural basis for a
spectral response, in which the parameter becomes momentum dependent,
\(\lambda\rightarrow\lambda(k)\).
\paragraph{Response at the amplitude level.}

We define the boundary-amplitude functional by
\begin{equation}
B_t^{(\lambda)}\psi_0
=\sqrt{\frac{\lambda\hbar}{m\pi}}\,
\partial_x\psi_{\mathrm{res}}(0,t).
\label{eq:Operetor B}
\end{equation} 
where the corresponding unidirectional current is
\begin{equation}
J_{\mathrm{UD}}(t;\psi_0)
=
\left|B_t^{(\lambda)}\psi_0\right|^2,
\end{equation}
The functional $B_t^{(\lambda)}$ is linear on its natural dense domain and
maps the initial state to $\mathbb C$. It is an amplitude-density functional,
not the positive arrival-time operator itself.

  Whereas the mathematical definition in Eq.~(\ref{eq:Operetor B}) is
straightforward, it also admits a natural physical interpretation.
One may regard \(\lambda\) as representing phenomenologically the interaction
of the wave function with the detection or absorption process taking place
at the boundary. Since this interaction acts at the level of the wave-function
amplitude, its corresponding weight enters as \(\sqrt{\lambda}\), while the
resulting detection rate is proportional to \(\lambda\).

 There is no reason, however, for the boundary interaction to affect all
spectral components in the same way. The response of a detector may depend on
the momentum, energy, or de Broglie wavelength of the incident component.
It is therefore natural to replace the constant parameter \(\lambda\) by a
spectral response function \(\lambda(k)\). In this interpretation,
\(\lambda(k)\) characterizes how the interaction taking place at the detector
boundary depends on the spectral properties of the incident quantum state.

We adopt a phenomenological detector model in which this response is taken
to be diagonal and phase-free in the single-channel spectral representation.
Each spectral component is therefore weighted independently by a
non-negative response \(\lambda(k)\).

The diagonal, phase-free form is adopted for simplicity. A more general
response may contain a \(k\)-dependent phase and off-diagonal couplings between
different spectral components. Such terms may modify the detailed time
dependence of the unidirectional current, but they do not affect the full-time
normalization condition used below to determine the diagonal spectral response
\(\lambda(k)\), as shown in Appendix B.

The corresponding boundary-amplitude functional is
\begin{equation}
B_t^{(\lambda)}\psi_0
=
\sqrt{\frac{\hbar}{m\pi}}
\int_{0}^{\infty}
\sqrt{\lambda(k)}\,
\widetilde\psi(k)\phi_k'(0)e^{-iE(k)t/\hbar}\,dk .
\label{eq:B_represent}
\end{equation}

Thus \(\sqrt{\lambda(k)}\) weights each spectral boundary amplitude before
the coherent sum is formed. Since \(\lambda(k)\ge0\), the generalized current
retains the positive form
\begin{equation}
J_{\mathrm{UD}}^{(\lambda)}(t;\psi_0)
=
\left|B_t^{(\lambda)}\psi_0\right|^2
=
\langle\psi_0,
B_t^{(\lambda)\dagger}B_t^{(\lambda)}
\psi_0\rangle
\ge0 .
\end{equation}

The remaining question is therefore not positivity, but normalization.
The condition under which this positive operator family defines a normalized
arrival-time POVM is derived in the following section.

%% file: sections/free_normalization.tex
\section{POVM Representation of the Unidirectional Current}
\label{sec:free_normalization}

Having established the positivity of the unidirectional current, we now
represent it as the expectation value of a positive operator. We define the
time-dependent current operator \(A_t\) through

\begin{equation}
J_{\mathrm{UD}}(t;\psi_0)
=
\langle \psi_0, A_t \psi_0\rangle .
\end{equation}

Using the boundary-amplitude functional \(B_t\)  (\ref{eq:B_represent}), the operator
is given by

\begin{equation*}
A_t = B_t^\dagger B_t .
\end{equation*}

It follows immediately that

\begin{equation*}
A_t \geq 0 ,
\end{equation*}

and therefore the unidirectional current is a positive quadratic form in the
initial state.

The remaining question is whether the positive operator family
\(A_t\,dt\) defines a normalized arrival-time POVM. This requires the
resolution-of-the-identity condition

\begin{equation}
\label{eq:Normalization_UC}
\int_{-\infty}^{\infty} A_t\,dt = I .
\end{equation}

 We now determine the condition under which this normalization is satisfied for a free particle on the negative half-line. This condition determines the admissible form of \(\lambda (k)\). The extension to non-free dynamics in the presence of a potential will be considered separately.

For the free particle on the negative half-line, the Dirichlet spectral
representation is labelled by \(k\geq0\), with dispersion relation

\begin{equation}
E(k)=\frac{\hbar^2k^2}{2m}.
\end{equation}

The restriction  \(k>0\) labels the continuous Dirichlet spectrum on the half-line and should not be interpreted as restricting the physical state to positive momentum; the threshold  \(k=0\) is understood through the \(\lim k\rightarrow 0^+\). Equivalently, the sine-transform representation corresponds to an odd extension to the full line, so that both positive- and negative-momentum Fourier components are present and related by the Dirichlet condition

\begin{equation*}
\widetilde\psi(-k)=-\widetilde\psi(k).
\end{equation*}

The expectation value of the unidirectional-current operator can then be
written as

\begin{equation}
\langle\psi,A_t^{(\lambda)}\psi\rangle
=
\frac{2\hbar}{\pi^2m}
\int_0^\infty dk
\int_0^\infty dk'\,
kk'\sqrt{\lambda(k)\lambda(k')}\,
\widetilde\psi(k)\widetilde\psi^*(k')\,
e^{-i(E(k)-E(k'))t/\hbar}.
\label{eq:At_expectation}
\end{equation}

The normalization requirement in Eq.~\eqref{eq:Normalization_UC} implies
that, for every normalized state,

\begin{equation*}
\int_{-\infty}^{\infty}dt\,
\langle\psi,A_t^{(\lambda)}\psi\rangle
=1.
\end{equation*}

Integrating Eq.~\eqref{eq:At_expectation} over the complete time axis gives

\begin{align*}
1
&=
\frac{2\hbar}{\pi^2m}
\int_0^\infty dk
\int_0^\infty dk'\,
kk'\sqrt{\lambda(k)\lambda(k')}\,
\widetilde\psi(k)\widetilde\psi^*(k')
\nonumber\\
&\qquad\times
\int_{-\infty}^{\infty}dt\,
e^{-i(E(k)-E(k'))t/\hbar}.
\end{align*}

Using

\begin{equation*}
\int_{-\infty}^{\infty}dt\,
e^{-i(E(k)-E(k'))t/\hbar}
=
2\pi\hbar\,\delta(E(k)-E(k')),
\end{equation*}

and, for \(k,k'>0\),

\begin{equation*}
\delta(E(k)-E(k'))
=
\frac{m}{\hbar^2k}\,
\delta(k-k'),
\end{equation*}

we obtain

\begin{equation*}
1
=
\int_0^\infty dk\,
\frac{4}{\pi}k\lambda(k)
|\widetilde\psi(k)|^2.
\end{equation*}

On the other hand, normalization of the state gives

\begin{equation*}
\int_0^\infty dk\,
|\widetilde\psi(k)|^2
=
1.
\end{equation*}

Since both relations must hold for every normalized state, it follows that

\begin{equation*}
\frac{4}{\pi}k\lambda(k)=1
\end{equation*}

for \(k>0\), and therefore

\begin{equation}
\boxed{
\lambda(k)=\frac{\pi}{4k}
}.
\label{eq:Lamda normalized}
\end{equation}
The selected response is therefore directly proportional to the de Broglie wavelength,

\[
\lambda(k)=\frac{1}{8}\lambda_\mathrm{bd}
\]

 The resulting response (\ref{eq:Lamda normalized}) defines a characteristic spatial scale associated with each spectral
component, increasing for longer de Broglie wavelengths and decreasing
for shorter ones.

The same response was originally identified by matching the MS boundary
amplitude to Kijowski's directional arrival-time amplitude. The independent
normalization argument given above shows that it is also selected directly
by the POVM normalization requirement. The original matching argument is
presented in Appendix~\ref{app:kijowski_lambda}.

The normalization requirement can also be relaxed to the subnormalization
condition
\begin{equation*}
\int_{-\infty}^{\infty} A_t^{(\lambda)}\,dt \leq I .
\end{equation*}
Which gives
\begin{equation}
0\leq\lambda(k)\leq\frac{\pi}{4k}
\qquad
\text{for almost every } k>0.
\end{equation}
Strict inequality on a set of nonzero spectral measure corresponds to
incomplete detection for states with spectral support in that set. We do
not consider the subnormalized case further here.

Thus, we have shown that the unidirectional-current operator family \(A_t\), with the response selected by the normalization condition, defines an arrival-time POVM. With positivity and normalization established, the operator family is now ready for a generalized Born-rule interpretation of the arrival-time statistics, distinct from the original MS hazard formulation of the unidirectional current. Since this construction does not introduce a self-adjoint time operator canonically conjugate to the Hamiltonian, Pauli's objection to such an operator does not apply.

\input{sections/finite_time}

%% file: sections/finite_time.tex
\paragraph{From complete-time arrival statistics to finite-time experiments.}
\label{subsec:finite_start}
 A natural interpretation is to regard the complete-time formulation as
an asymptotic idealization, analogous to the use of incoming asymptotic
states in scattering theory~\cite{Shankar1994}. In this interpretation,
the state prepared at \(t=0\) is the normalized surviving subensemble of
an ensemble evolved from the remote past and conditioned on no arrival
before \(t=0\). If
\begin{equation}
P_{<0}
=
\int_{-\infty}^{0}
\left\langle
\psi_0,
A_t^{(\lambda)}\psi_0
\right\rangle\,dt .
\end{equation}
then the survival probability at the preparation time is
\begin{equation*}
P_{\mathrm{surv}}(0)=1-P_{<0}.
\end{equation*}
The finite-start state is therefore obtained by the normalization
\begin{equation}
|\psi_0\rangle
=
\frac{|\psi_{\mathrm{surv}}(0)\rangle}
{\sqrt{P_{\mathrm{surv}}(0)}} .
\end{equation}
The arrival distribution measured after preparation is then
\begin{equation*}
p(t;\psi_0)
=
\langle\psi_0,A_t^{(\lambda)}\psi_0\rangle ,
\qquad t\geq0 .
\end{equation*}

This interpretation does not imply that the laboratory particle was
literally prepared at \(t=-\infty\); rather, it provides a connection
between the complete-time normalization and a state prepared at a finite
laboratory time.

%% file: sections/statistical_interpretations.tex
 \subsection{From the MS Hazard Law to a Generalized Born-Rule Interpretation}
\label{subsec:POVM_interpretation}

As introduced in the original MS formulation and discussed above, the
unidirectional current was interpreted as an instantaneous hazard for first
arrival~\cite{MarchewkaSchuss1998,MarchewkaSchuss2000,
MarchewkaSchuss2002}. We now introduce an alternative interpretation of the
unidirectional current in terms of a POVM, placing it directly within the
quantum-mechanical measurement framework through the generalized Born rule.

For each fixed time \(t\), the current is represented by the positive
self-adjoint operator
\begin{equation}
A_t^{(\lambda)}
=
A_t^{(\lambda)\dagger}
\geq 0,
\qquad
J_{\mathrm{UD}}^{(\lambda)}(t;\psi_0)
=
\langle\psi_0,A_t^{(\lambda)}\psi_0\rangle .
\end{equation}
However, because the Hamiltonian is bounded from below, the usual
construction of a self-adjoint time operator canonically conjugate to the
Hamiltonian is obstructed; in the time-of-arrival problem this naturally
leads to a POVM description \cite{EgusquizaMuga1999}.
Arrival time is therefore described through the generalized Born rule.

For any time interval \(\Delta\), we define
\begin{equation*}
E(\Delta)
=
\int_\Delta A_t^{(\lambda)}\,dt,
\qquad
E(\mathbb{R})=I .
\end{equation*}
The generalized Born rule then assigns the probability
\begin{equation*}
\Pr\{T\in\Delta\mid\psi_0\}
=
\langle\psi_0,E(\Delta)\psi_0\rangle .
\end{equation*}

For an arbitrarily small interval \([t,t+dt]\),
\begin{equation}
\Pr\{T\in[t,t+dt]\mid\psi_0\}
=
\langle\psi_0,A_t^{(\lambda)}\psi_0\rangle\,dt
=
J_{\mathrm{UD}}^{(\lambda)}(t;\psi_0)\,dt .
\end{equation}
 Thus, the generalized Born rule identifies the normalized unidirectional current with the arrival-time probability density

The distinction between the hazard and POVM interpretations of the
unidirectional current is therefore substantive. In the original MS
formulation, the current enters as a conditional hazard rate, and the
first-arrival statistics are obtained through a survival process analogous
in structure to a classical rate equation. In the present formulation, by
contrast, the normalized current is directly the density of an arrival-time
POVM and therefore enters the generalized Born rule without an additional
survival factor. The POVM formulation thus places the unidirectional current
directly within the quantum-mechanical measurement framework. Which of these
two statistical interpretations provides the appropriate physical
description is ultimately an empirical question to be decided by experiment.

%% file: sections/discussion.tex
\section{Conclusion}
\label{sec:conclusion}

We have shown that the MS unidirectional first-arrival current admits a
positive-operator representation and, on this basis, developed a generalized
Born-rule interpretation of the MS construction. The detector response was
introduced at the amplitude level through a non-negative spectral function
\(\lambda(k)\), representing the effective particle--detector interaction.
Complete-time normalization then uniquely selects
\[
\lambda(k)=\frac{\pi}{4k},
\]
thereby promoting the positive-operator construction to a normalized
arrival-time POVM. We also discussed possible physical realizations of this
spectral response.

Kijowski's arrival-time distribution provides a natural reference point
for comparison, due to its important role as a mathematically consistent
POVM construction and its central position in the theory of quantum
arrival times. Its comparison with the MS construction highlights both
the equivalence of the resulting arrival-time statistics and the
different physical interpretations underlying the two approaches.
For the normalized response, we established an exact MS--Kijowski identity:
each one-sided MS unidirectional current is related to the corresponding
directional branch of Kijowski's arrival-time distribution. This identity provides, within this framework, a physical interpretation of
the directional Kijowski branches in terms of one-sided first-arrival
processes and a mathematical
tool for relating the simple half-line MS spectral structure to properties
of the Kijowski distribution, including its long-time behavior and the
existence of arrival-time moments.

The resulting framework therefore provides a dynamical first-arrival process
with a positive-operator and POVM measurement structure, a spectral
interpretation of the detector response, and a direct connection between
the MS first-arrival construction and Kijowski's arrival-time law.
 Within this framework, the original MS hazard interpretation and the
generalized Born-rule interpretation remain statistically distinct and can,
in principle, be distinguished experimentally.

\section*{Acknowledgments}

I am grateful to Will Cavendish for fruitful discussions on the quantum
time-of-arrival problem and for his comments and feedback on the manuscript.
In particular, his comments motivated the independent normalization analysis
leading to the spectral response $\lambda(k)=\pi/(4k)$.

%% file: sections/appendix_kijowski.tex
  
\section{Alternative Determination of the Spectral Response from the
MS--Kijowski Correspondence}
\label{app:kijowski_lambda}

The spectral response fixed by the normalization condition is independently
recovered from the MS--Kijowski correspondence. This provides an alternative
derivation of Eq.~(11). Thus, the same spectral response emerges from two independent structures:
the resolution-of-the-identity condition and the MS--Kijowski correspondence. 

For the free particle on the left half-line, the normalized Dirichlet
eigenfunctions are
\begin{equation*}
\phi_k^L(x)=\sqrt{\frac{2}{\pi}}\sin(kx),
\qquad k>0,
\end{equation*}
so that
\begin{equation*}
\left.\partial_x\phi_k^L(x)\right|_{x=0}
=
\sqrt{\frac{2}{\pi}}\,k .
\end{equation*}

Substitution into the general boundary-amplitude functional gives
\begin{equation*}
B_{L,t}^{(\lambda)}\psi_L
=
\frac{\sqrt{2\hbar/m}}{\pi}
\int_0^\infty
k\sqrt{\lambda(k)}\,
\widetilde{\psi}_L(k)
e^{-i\hbar k^2t/(2m)}\,dk .
\label{eq:appendix_MS_amplitude}
\end{equation*}

Kijowski's positive-momentum amplitude is
\begin{equation*}
A_+(t)
=
\sqrt{\frac{\hbar}{2\pi m}}
\int_0^\infty
\sqrt{k}\,
\widetilde{\psi}_+(k)
e^{-i\hbar k^2t/(2m)}\,dk .
\label{eq:appendix_K_amplitude}
\end{equation*}

As shown in Sec.~5, the normalized full-line extension of the half-line
state satisfies
\begin{equation*}
\widetilde{\psi}_+(k)
=
\frac{1}{\sqrt{2}}\widetilde{\psi}_L(k),
\qquad k>0,
\end{equation*}
and the corresponding amplitudes obey
\begin{equation*}
B_{L,t}^{(\lambda)}\psi_L
=
i\sqrt{2}\,A_+(t).
\end{equation*}

Since this relation must hold for arbitrary spectral states, comparison of
the integrands gives, up to the irrelevant phase,
\begin{equation*}
\frac{\sqrt{2\hbar/m}}{\pi}\,
k\sqrt{\lambda(k)}
=
\sqrt{\frac{\hbar}{2\pi m}}\sqrt{k}.
\end{equation*}
Hence

\begin{equation*}
\boxed{\lambda(k)=\frac{\pi}{4k}}.
\end{equation*}

Thus the MS--Kijowski correspondence independently selects exactly the
same spectral response as the resolution-of-the-identity condition
derived in Sec.~4.
 
\section{General Spectral Structure and Full-Time Normalization}

To verify that the simplifications adopted in the main text do not affect the
normalization result, consider a general spectral contribution to the current
of the form

\begin{equation*}
J(t)
=
\int_0^\infty dk
\int_0^\infty dk'\,
F(k,k')\,
e^{i[\theta(k)-\theta(k')]}
e^{-i[E(k)-E(k')]t/\hbar},
\end{equation*}

where \(F(k,k')\) is an arbitrary spectral kernel and
\(\theta(k)\) is an arbitrary \(k\)-dependent phase.

Integrating over the full time axis gives

\begin{equation*}
\int_{-\infty}^{\infty} dt\, J(t)
=
2\pi\hbar
\int_0^\infty dk
\int_0^\infty dk'\,
F(k,k')\,
e^{i[\theta(k)-\theta(k')]}
\delta(E(k)-E(k')).
\end{equation*}

For the free-particle dispersion relation

\begin{equation*}
E(k)=\frac{\hbar^2k^2}{2m},
\end{equation*}

and in the one-directional sector \(k,k'>0\),

\begin{equation*}
\delta(E(k)-E(k'))
=
\frac{m}{\hbar^2 k}\,
\delta(k-k').
\end{equation*}

The \(k'\) integration therefore reduces the result to

\begin{equation*}
\int_{-\infty}^{\infty} dt\, J(t)
=
\frac{2\pi m}{\hbar}
\int_0^\infty
\frac{dk}{k}\,
F(k,k).
\end{equation*}

The phase factor has disappeared because

\begin{equation*}
e^{i[\theta(k)-\theta(k)]}=1,
\end{equation*}

and only the diagonal part \(F(k,k)\) contributes to the full-time
normalization.

Thus, possible \(k\)-dependent phase factors and off-diagonal spectral
structure may modify the detailed time dependence of the current, but they do
not affect the full-time normalization result. The simplifications adopted in
the main text therefore leave unchanged the normalization condition used to
determine the diagonal spectral response.

\appendix